# Effective quantum number for describing nanosystems

N.N.Trunov

*D.I.Mendeleyev Institute for Metrology*
*Russia, St.Peterburg, 19005 Moskovsky pr. 19*

**Abstract** An enormous variety of quantum nanoobjects and nanosystems calls for the development of new approaches to their description and parametrization. Corresponding methods should be simple, universal enough and ensuring the retention of essential part of information at small number of parameters.

A universal effective quantum number was introduced earlier and successfully used for describing centrally symmetric systems. But in most cases nanosystems own a lower symmetry. In the present article we generalize and adapt this method for such systems.

## 1. Introduction

The number of known nanoobjects and nanosystems is vast and continues to increase, while their properties are essentially infinite in variety. A complete, rigorous theoretical description of all objects of this kind is both practically unrealizable and, in many cases, also inappropriate because of its complexity and excessive detail. Thus, efficient parametrization is needed for each type of nanosystem, in order to provide the essential and secondary information on a given object. This kind of approach requires the development and adaptation of various physical and mathematical methods depending on the objects themselves and the tasks for which they are intended: various asymptotic and approximation techniques, reduced description etc [1,2].

In particular, it was shown that centrally symmetric multiparticle systems such as clusters may be simply and successfully described by means of the effective quantum number [3, 4]

$$T = (n + 1/2) + \varphi(l + 1/2) = T(\nu, \lambda) \qquad (1)$$

with $n$ and $l$ being the radial and orbital quantum numbers respectively. The value of $T$ determines the level-ordering and values of the bound-state energies satisfy with a good accuracy from the following modified semiclassical condition

$$\Phi(E) = \frac{\sqrt{2m}}{\pi \hbar} \int \sqrt{E - V} \, dr = T \qquad (2)$$

where $V(r)$ is a given potential without the centrifugal one. The main parameter $\varphi$ may be calculated i.e. as

$$\varphi^2 = \frac{1}{2\pi^2} \frac{\left[\int dr \sqrt{2(E-V)}\right]^3}{\int r^2 dr \left[2(E-V)\right]^{3/2}} \qquad (3)$$

or found from experimental dates.

However, many nanosystems have a lower symmetry than the central one. Thus a new problem arises: how to generalize our *T*-method for such cases.

## 2. An elliptic deformation

Suppose we have a small change of a given potential

$$V\left(\frac{x^2+y^2+z^2}{R^2}\right) \to V\left(\frac{x^2+y^2}{a^2}+\frac{z^2}{c^2}\right), \qquad (4)$$

such that the "characteristic volume" remains unchanged:

$$R^3 = a^2c = const, \quad a = R(1-\alpha/3), \quad c = R(1+2\alpha/3), \quad |\alpha|<<1. \quad (5)$$

Hereafter we show that in this case we become

$$T(n,l,m) = [\ 1+\alpha f(m,l)/2\ ]\ T(n,l), \qquad (6)$$

instead of the previous $T(n,l)$ so that a dependence on the magnetic quantum number $m$ appears,

$$f(m,l) = \frac{1}{(2l-1)(2l+3)}\left[m^2 - \frac{l(l+1)}{3}\right]. \qquad (7)$$

In order to prove (6), introducing new variables

$$\tilde{x}, \tilde{y} = \frac{R}{a}(x,y); \qquad \tilde{z} = \frac{R}{C}z, \qquad (8)$$

we return to the previous centrally symmetric potential, but the operator of the kinetic energy acquires an addition

$$H' = -\frac{\alpha}{3}\left(\Delta - 3\frac{\partial^2}{\partial z^2}\right). \qquad (9)$$

The corresponding addition $\Delta E$ to the energy of the bound state $E\ (n,l,)$ was found earlier [5, sect. 38] for

$$V = 0, \quad r < R, \qquad (10)$$
$$V = \infty, \quad r \geq R.$$

In this special case

$$\Delta E(n,l,m) = 4\alpha E(n,l)f(m,l) \qquad (11)$$

with $f(m,l)$ from (7). By means of analogous calculations we get for an arbitrary $V(r)$ instead of (11)

$$\Delta E(n,l,m) = 4\alpha K(n,l)f(m,l) \qquad (12)$$

with $K$ being the mean kinetic energy ( $E = K$ for (10)).

If we have a power-law potential $r^\beta$, the virial theorem states that

$$K = \beta E/(\beta + 2). \qquad (13)$$

It is known [6] that in such (undisturbed) potentials

$$E(n,l) = F(\beta)T(n,l)^{\frac{2\beta}{2+\beta}} \qquad (14)$$

with a constant $F$, so that an addition to $T$

$$\frac{\Delta E}{E} = \frac{2\beta}{2+\beta}\frac{\Delta T}{T}, \qquad T = \nu + \psi\lambda. \qquad (15)$$

Combining (12), (13) and (15) we found

$$\Delta T(n,l,m) = \frac{\alpha}{2}f(m,l)T(n,l) \qquad (16)$$

and finally (6). It is remarkable the disappearance of $\beta$, so we expect that (6) is valid for arbitrary potentials, at least power-like ones.

## 3. Effective quantum number for quantum dots

It is supposed [7], that so called quantum dots may be represented by the following potential in cylindrical coordinates $(z, \rho)$:

$$V(z, \rho) = v(\rho) \qquad 0 < z < d, \qquad (17)$$
$$V(z, \rho) = \infty \qquad z < 0, z > d.$$

Thus energies of bound states are divided into two parts:

$$E(N, n, m) = K(N) + E(n, m), \qquad (18)$$

where $K$ – the known energy of the free motion along z-axis, $N = 1, 2…$ In many cases only states with the lowest $N = 0$ are actual [7].

$E(n,m)$ corresponds to two-dimensional bound states with $v(r)$. In a general case of an arbitrary dimensionality $D$ the effective number $T$ includes

$$\lambda = l + (D - 2)/2. \qquad (19)$$

Now we have $D = 2$ so that $\lambda = l = |m,|$ where we introduce the usual notation $m$ for the angular number.

For each fixed N the level-ordering obeys the increasing number

$$T = (n + ½) + \varphi |m|. \qquad (20)$$

Then (2) and (3) are still valid with $v(r)$ instead of $V(r)$.

A universal diagram for bound states as functions of $\varphi$ as well as its values for actual potentials are given in [3]. In the case of power-law potentials we have two exact values: $\varphi = 1$ for the Coulomb potential, $\beta = -1$ and $\varphi = ½$ for the oscillator, $\beta = 2$. In the limiting case of an empty cavity with impenetrable boundaries $\beta = \infty$ and $\varphi \approx 0.39$.

It should be stressed that different potentials may have coinciding values of $\varphi$ and $T$. That is why namely they are effective reduced parameters.

1. V.S.Aleksandrov, N.N.Trunov and A.A.Lobashev. Measur. Techn., **55**, No 7, 763-768 (2012).
2. N.N.Trunov. On new approaches to the study of quantum nanosystems. arXiv: 1302. 6016 [physics. gen-ph] (2013).
3. Lobashev A. A., Trunov N. N. A universal effective quantum number for centrally symmetric problems // J. Phys. A: Math. Theor. – 2009. - V. 42. – P. 345202.
4. N.N.Trunov. Renormalized effective quantum number for centrally symmetric problems. arXiv: 1002. 4771 [math-ph] (2010).
5. L.D.Landau and E.M.Lishitz. Quantum mechanics. Engl. transl., Oxford Univ. Press, Oxford (1975).
6. A.A.Lobashev and N.N.Trunov. An integral semiclassical method for calculating the spectra for centrally symmetric potentials. Theor. Math. Phys. (Engl. Transl.), **120**, 1250-264 (2000).
7. E.Roduner. Nanoscopic materials. Size-dependent phenomena. RSCPublishing, Lond. (2006).